\documentstyle[aps,epsfig,floats,preprint]{revtex}
\begin{document}
\draft 
\preprint{\vbox{\hbox{{\tt [ SOGANG-HEP 291/01 | hep-th/0201117 ]}}}}
\title{Scattering amplitudes and thermal temperatures of the Schwarzschild-de Sitter black holes}
\author
{Won Tae Kim\footnote{electronic address:wtkim@ccs.sogang.ac.kr}, John
  J. Oh\footnote{electronic address:john5@string.sogang.ac.kr}, and
  Ki Hyuk Yee\footnote{electronic address:quicksilver@string.sogang.ac.kr}}
\address{Department of Physics and Basic Science Research Institute,\\
         Sogang University, C.P.O. Box 1142, Seoul 100-611, Korea}
\date{\today}
\maketitle
\begin{abstract}
We study thermodynamic evaporation of Schwarzschild-de Sitter black
holes in terms of a low energy perturbation
theory. A small black hole which is far from the cosmological horizon
and observers at the spacelike hypersurface where black hole
attraction and expansion of cosmological horizon balance exactly are
considered. In the low energy perturbation, scalar field equations are solved
in both regions of the hypersurface and scattering amplitudes are
derived. And then the desired thermal temperatures from the two
horizons are obtained as a ``minimal'' value of the statistical
thermal temperature, and the fine-tuning between amplitudes gives a
relation of the two temperatures.
\end{abstract}
\bigskip
 
\section{Introduction\hfil{}}
Recent phenomenological observations show that the expansion of
our universe is accelerating \cite{per}, which implies that our
universe will look like de Sitter(dS) spacetimes that have
energies dominated by a negative pressure such as cosmological
constant or quintessence. A model of
black holes corresponding to this universe can be considered, which is
described by the Einstein-Hilbert action with a positive cosmological
constant, and the static and neutral solution in this system has been
known as a Schwarzschild-de Sitter(SSdS) black hole. In this family of black
holes, the size of black hole varies from zero to its cosmological
limitation and then the spacetimes where we live are confined within
the cosmological horizon.  

Recently, Hawking et. al. have studied black holes
with the cosmological horizon in Refs. \cite{bh1,bh2,gh}. If the size
of the black hole is much smaller than that of the cosmological horizon, the
radiation coming from the cosmological horizon is negligible compared
to that from the black hole horizon, and the effective radiation 
experienced by observers is similar to the case of Schwarzschild(SS) black
holes since the thermal temperature from black hole horizon $T_{\rm
  BH}$ is greater than that from cosmological horizon $T_{\rm DS}$. So
the black hole will radiate more than what it receives, and there
exists an energy transfer from the black hole horizon $r_{h}$ to the
cosmological horizon $r_{c}$. Finally, the black hole horizon becomes
to shrink completely while the cosmological horizon grows faster until
the black hole disappears to evaporate out. Furthermore, the thermodynamic
instability for this case has been investigated in Ref. \cite{gp}. 
For the black hole which grows up to the same size of the cosmological
horizon, the radiation coming from cosmological horizon equals to the
black hole radiation, and it will be in the thermal equilibrium. For
the nearly degenerate case, if the black hole horizon is shrunk from
its equilibrium value, the size of black hole increases by
``anti-evaporation'' while the black hole evaporates when we choose a
different type of initial perturbations which corresponds to a kind of
``push'' in the radiation bath.

On the other hand, it would be interesting to study scattering
amplitudes of the scalar field on the SSdS black hole background since the
greybody factor or the decay rate is closely related to
the thermal temperature of black holes. Some studies on these
quantities in terms of the low energy perturbation method have been
done for the four-dimensional Kerr-Newman(KN) black holes \cite{ms,cl}
and the Kerr-Newman-de Sitter(KN-dS) black hole without considering
observers \cite{stu}. 

In this paper, we shall study on the scattering amplitudes of the
massive scalar field on the
SSdS black hole background by using the low energy perturbation theory. We
consider a small black hole whose size is much smaller than that of
the cosmological horizon and assume that the energy of probing fields
$\omega$, is negligible compared to the inverse of each horizon. In
this configuration, the whole spacetimes can be approximately divided
by two parts - SS black hole and dS spacetimes, and then the massive
scalar field equation is easily solved on each background. Finally,
the scattering problem in SSdS background can be solved by matching
coefficients between asymptotic and near horizon solutions, and
defining reflection coefficients. In addition, the temperatures from
the two horizons can be obtained as ``minimal'' values of statistical
thermal temperatures derived from the reflection coefficients.
In Sec. II, we present a model of Einstein-Hilbert action with the
negative cosmological constant, which shows two characteristic
properties - SS black hole and dS spacetimes, and set up the static
configuration of this model. Section III and IV contain the specific
calculation of low energy perturbation in the attraction dominant
region and the expansion dominant region. As a result, the expected
thermal temperatures at both regions are obtained by using the
appropriate boundary condition and coefficient matching procedure. 
In Sec. V, some discussions on the fine-tuning of amplitudes between
different two regions and thermodynamic behaviors of non-degenerate and
degenerate cases are briefly presented.

\section{Model Setup \hfil{}}
We start with a four-dimensional Einstein-Hilbert action with a
cosmological constant $\Lambda > 0$, which is given by
\begin{equation}
  \label{action}
  S_{\rm EH}^{(\Lambda)} = \frac{1}{16\pi G} \int d^{4}x \sqrt{-g}\left(R-\Lambda\right),
\end{equation}
where $G$ is a gravitational constant and we set $G=1$. From
Eq. (\ref{action}), the neutral and static solution for the
spherically symmetric Einstein equation is described by the
SSdS black hole metric
\begin{equation}
  \label{metric}
  (ds)^2 = - U(r)dt^2 + U(r)^{-1} dr^2 + r^2 d\Omega_{(2)}^2,
\end{equation}
where
\begin{equation}
  \label{u(r)}
  U(r) = 1-\frac{2M}{r} - \frac{\Lambda}{3}r^2,
\end{equation}
and $\Omega_{(2)}^2$ is a unit line element of two-sphere, and $M$ is
a black hole mass. Note that two horizons $r_{h}$ and $r_{c}$ ($r_{h}
< r_{c}$) are only possible for $0 < M < (3\sqrt{\Lambda})^{-1}$. In
this spacetime, there exits a two-spherical hypersurface where the
gravitational attraction and cosmological expansion balance out
exactly, so a geodesic orbit is defined as the position where
observers need no acceleration in order to stay. From
Eq. (\ref{metric}), the geodesic orbit is located at $r_{g} =
(3M\Lambda^{-1})^{1/3}$ with a normalized Killing vector at this
hypersurface. Note that it is similar to the case of an observer who
lives at infinity for the SS black hole. 

\begin{figure}[htbp]
    \begin{center}
    \leavevmode
    \centerline{
        \epsfig{figure=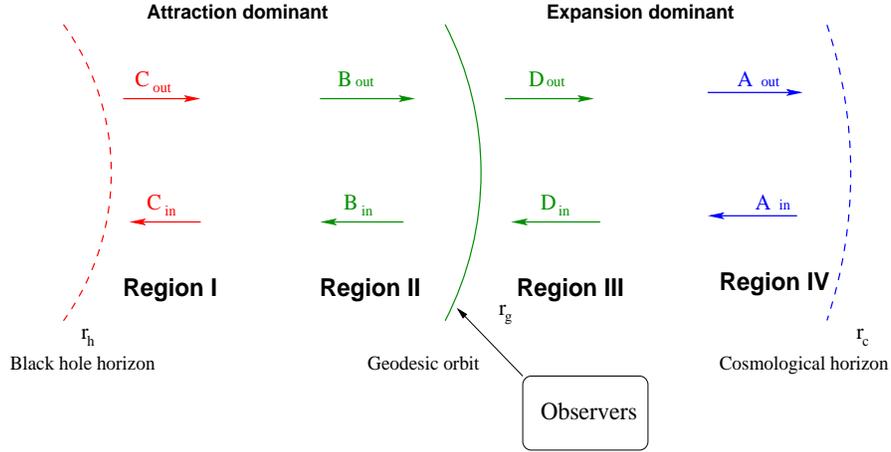, width=12cm, height=6cm}
        }
    \caption{Configuration of Schwarzschild-de Sitter black holes:
    {\it Observers are located at the two-spherical spacelike
    hypersurface where cosmological expansion and black hole
    attraction balance out exactly so that there is no acceleration on
    this hypersurface as like observers at Minkowski spacetimes, and
    as for a thermal radiation, observers will experience thermal
    temperatures $T_{\rm BH}$ and $T_{\rm DS}$ coming from two horizons.}} 
    \label{fig:ssds}
    \end{center}
\end{figure}

We can divide our whole spacetimes into specific four regions as shown in the
Fig. (\ref{fig:ssds}). With the assumption of $r_{h}<<r_{g}<<r_{c}$,
as $r$ goes to the black hole horizon $r_{h}$, the SS black
hole metric is prominent in the attraction dominant region (region (I)
and (II)) in as 
\begin{equation}
  \label{ss}
  U(r) \approx 1-\frac{2M}{r},
\end{equation}
while if we are in the expansion dominant region (region (III) and
(IV)) which is accomplished by $r\rightarrow r_{c}>>r_{g}$, the metric
(\ref{u(r)}) behaves as like a dS metric
\begin{equation}
  \label{ds}
  U(r) \approx 1-\frac{\Lambda}{3}r^2.
\end{equation}
Note that the metric element (\ref{u(r)}) is
a patch of Eqs. (\ref{ss}) and (\ref{ds}) at the geodesic orbit
$r_{g}$ but this picture is only valid for $r_{h} << r_{g} << r_{c}$.
For each region, a field equation can be solved and the corresponding
amplitudes of solutions between the four regions will be matched
exactly. Then we define a reflection coefficient as a ratio between
ingoing and outgoing fluxes in region (II) and region (III), and
finally obtain the scattering amplitudes and thermal temperatures
coming from these two horizons. 

The dynamical behavior of a massive scalar field under the background
(\ref{metric}) is described by
\begin{equation}
  \label{kgeqn}
  \left(\Box - m^2 \right) \Psi(t,r,\theta, \phi)=0. 
\end{equation}
Using the separation of variables, $\Psi(t,r,\theta, \phi) =
e^{-i\omega t}R(r)\Theta(\theta) e^{i\mu\phi}$, the field equation
(\ref{kgeqn}) becomes
\begin{eqnarray}
  & &\frac{1}{r^2} \partial_{r}\left(r^2 U(r) \partial_{r} R(r)\right) +
  \left(\frac{\omega^2}{U(r)} - \frac{\nu(\nu+1)}{r^2} - m^2 \right)
  R(r) = 0, \label{radial}\\
  & &\frac{1}{\sin\theta} \partial_{\theta} \left(\sin\theta
  \partial_{\theta} \Theta(\theta)\right) -
  \frac{\mu^2}{\sin^2\theta}\Theta(\theta) = -
  \nu(\nu+1)\Theta(\theta), \label{spherical}
\end{eqnarray}
where $\nu$ is an angular mode.

\section{Low Energy Perturbation in the Black Hole Region}
\subsection{Solution near the black holes(Region (I) and (II))}
At first, let us consider the field equation (\ref{radial}) in the near
black hole horizon region(Region (I) in Fig. (\ref{fig:ssds})). As $r$ goes to
$r_{h}$, the metric (\ref{metric}) becomes to be a SS metric, $U(r) \approx (1-2M/r)$. So the equation of motion
(\ref{radial}) can be written as
\begin{equation}
  \label{sseqn}
  \left(1-\frac{r_{h}}{r}\right)\partial_{r}^2 R(r) +
  \left(\frac{2}{r} - \frac{r_{h}}{r^2}\right)\partial_{r}R(r)
  +\left(\frac{\omega^2}{\left(1-\frac{r_{h}}{r}\right)} -
  \frac{\nu(\nu+1)}{r^2} - m^2\right)R(r) = 0,
\end{equation}
where the horizon $r_{h} \approx 2M$ as $r \rightarrow r_{h}$.
Taking a coordinate transformation such as $z=(r-r_{h})/r$, where $z$
spans from $0$ to $1$ as $r$ goes from $r_{h}$ to $r_{g}$, where we
assumed that $r_{h} << r_{g}$, the radial equation becomes
\begin{equation}
  \label{ztranseqn}
  z(1-z) \partial_{z}^2 R(z) + (1-z) \partial_{z}R(z) +
  \left(\frac{\omega^2 r_{h}^2}{z} - \frac{\nu(\nu+1)}{1-z} - m^2
  r_{h}^2\right)R(z) = 0.
\end{equation}
Using $R(z) = z^{\alpha}(1-z)^{\beta}g(z)$ in order to remove two
singular points at $z=0$ and $z=1$, Eq. (\ref{ztranseqn}) is
written as
\begin{eqnarray}
  \label{zint}
   & &z(1-z)\partial_{z}^2 g(z) + (1+2\alpha -
  (1+2\alpha+2\beta)z)\partial_{z}g(z) \nonumber \\ & &+\frac{1}{z}(\alpha^2 +
  \omega^2 r_{h}^2)g(z) + \frac{1}{1-z}(\beta^2 -\beta
  -\nu(\nu+1))g(z) - ((\alpha+\beta)^2 + m^2
  r_{h}^2)g(z) = 0,
\end{eqnarray}
and we have $\alpha = \pm i\omega r_{h}$ and $\beta = \nu +1$. Then
Eq. (\ref{zint}) becomes 
\begin{equation}
  \label{zeqn}
  z(1-z)\partial_{z}^2 g(z) + (1+2\alpha -
  (1+2\alpha+2\beta)z)\partial_{z}g(z) - ((\alpha+\beta)^2 + m^2
  r_{h}^2)g(z) = 0.
\end{equation}
Note that we take the plus signature of $\alpha$ since the solution
will be symmetric for changing $\alpha$ into $-\alpha$.
The solution of Eq. (\ref{zeqn}) is given as
\begin{eqnarray}
  \label{sssol}
  R_{\rm (I) } (r) =& & C_{\rm out} z^{\alpha} (1-z)^{\beta} F(\alpha_{-} +
  \beta, \alpha_{+} + \beta, 1+2\alpha ;z)\nonumber \\& &
  \qquad + C_{\rm in} z^{-\alpha}
  (1-z)^{\beta} F(-\alpha_{+} +
  \beta, -\alpha_{-} + \beta, 1-2\alpha ;z),
\end{eqnarray}
where $\alpha_{\pm} = \alpha \pm imr_{h} = i(\omega\pm m)r_{h}$ and
$C_{\rm in}$ and $C_{\rm out}$ are ingoing and outgoing coefficients,
respectively. 

Next, from the solution (\ref{sssol}), we shall use a $z \rightarrow
1-z$ transformation \cite{as}, 
\begin{eqnarray}
  \label{z1ztrans1}
  F(a,b;a+b-l;z) &=& \frac{\Gamma(l)\Gamma(a+b-l)}{\Gamma(a)\Gamma(b)}
  (1-z)^{-l} \sum_{n=0}^{l-1} \frac{(a-l)_{n}(b-l)_{n}}{n! (1-l)_{n}}
  (1-z)^{n} \nonumber \\
  &-& \frac{(-1)^{l} \Gamma(a+b-l)}{\Gamma(a-l)\Gamma(b-l)}
  \sum_{n=0}^{\infty} \frac{(a)_{n}
  (b)_{n}}{n!(n+l)!}(1-z)^{n}\nonumber \\
  &\times& \left[\ln (1-z) - \psi(n+1) - \psi(n+l+1)
  + \psi(a+n) + \psi(b+n)\right],
\end{eqnarray}
where $\psi(z)$ is a digamma function and $a = \pm
\alpha_{\mp}+\beta$, $b = \pm \alpha_{\pm}+\beta$, and $l=2\nu+1$ in our case.
As $z\rightarrow 1$, for the lowest order of $n$, the solution
(\ref{sssol}) is transformed to the region (II) by using
Eq. (\ref{z1ztrans1}) as
\begin{eqnarray}
  \label{z1sol}
  R_{z\rightarrow 1} (r) &=&  \left[ C_{\rm out} \frac{\Gamma(2\beta -
  1)\Gamma(1+2\alpha)}{\Gamma(\alpha_{-}+\beta)\Gamma(\alpha_{+}+\beta)} + C_{\rm in}\frac{\Gamma(2\beta -
  1)\Gamma(1-2\alpha)}{\Gamma(-\alpha_{+}+\beta)\Gamma(-\alpha_{-}+\beta)}\right]\left(\frac{r_{h}}{r} \right)^{1-\beta} \nonumber \\
  &+& \left[ C_{\rm out} \frac{{\cal D}_{\nu}^{\rm
        out}\Gamma(1+2\alpha)}{\Gamma(1+\alpha_{-}-\beta)\Gamma(1+\alpha_{+}-\beta)} + C_{\rm in} \frac{{\cal D}_{\nu}^{\rm in}\Gamma(1-2\alpha)}{\Gamma(1-\alpha_{+}-\beta)\Gamma(1-\alpha_{-}-\beta)}\right]\left(\frac{r_{h}}{r} \right)^{\beta},
\end{eqnarray}
where ${\cal D}_{\nu}^{\rm in} = -\psi(1) -\psi(2\beta) +
\psi(-\alpha_{+}+\beta) + \psi(-\alpha_{-}+\beta) = {\cal
  D}_{\nu}^{\ast \rm out}$ and note that Eq. (\ref{z1sol}) is
symmetric for changing $\beta$ into $1-\beta$.

\subsection{Asymptotic solution at Region (II)}
We now consider the asymptotic limit of $r \rightarrow r_{g} >> r_{h}$
which describes the region (II) in Fig. (\ref{fig:ssds}). In
this limit, the field equation (\ref{sseqn}) becomes
\begin{equation}
  \label{asymss}
  \partial_{r}^2 R(r) + \frac{2}{r} \partial_{r} R(r) + \left(p^2 -
  \frac{\nu(\nu+1)}{r^2}\right)R(r) = 0,
\end{equation}
where $p = \sqrt{\omega^2 - m^2}$, and the solution is given as Bessel
functions,
\begin{equation}
  \label{bessel}
  R_{r_{g}}^{(z)} (r) = \frac{B_{1}}{\sqrt{r}}
  J_{-\frac{1}{2}(1-2\beta)}(pr) + \frac{B_{2}}{\sqrt{r}} J_{\frac{1}{2}(1-2\beta)}(pr),
\end{equation}
where $B_{1}$ and $B_{2}$ are amplitudes of the solution.
The solution (\ref{bessel}) can be expanded by keeping the lowest
leading order of $r$ and defining new coefficients $B_{\rm in}$ and
$B_{\rm out}$ as
\begin{eqnarray}
  \label{defcoef1}
  & & B_{1} = B_{\rm in} + B_{\rm out} \nonumber \\
  & & B_{2} = i(B_{\rm in} - B_{\rm out}),
\end{eqnarray}
in order to separate wave modes into in-and out-going parts,
the solution is written as
\begin{equation}
  \label{solasym}
  R_{r_{g}}^{(z)}(r) = B_{\rm in} \left( \Omega_{\beta}^{(+)}
  \frac{1}{r^{1-\beta}} + i\Omega_{\beta}^{(-)} \frac{1}{r^{\beta}}\right)+B_{\rm out} \left( \Omega_{\beta}^{(+)}
  \frac{1}{r^{1-\beta}} - i\Omega_{\beta}^{(-)} \frac{1}{r^{\beta}}\right),
\end{equation}
where $\Omega_{\beta}^{(+)} \equiv \left(\frac{1}{2}p\right)^{-\frac{1}{2}(1-2\beta)}/{\Gamma\left(\frac{1}{2} +
  \beta\right)}$ and
  $\Omega_{\beta}^{(-)} \equiv \left(\frac{1}{2}p\right)^{\frac{1}{2}(1-2\beta)}/{\Gamma\left(\frac{3}{2} -
  \beta\right)}$. 
Note that $\Omega_{\beta}^{(\pm)\ast} = \Omega_{\beta}^{(\pm)}$ and
$\Omega_{\beta}^{(+)} \Omega_{\beta}^{(-)}$ is explicitly calculated as $(\Omega_{\beta}^{(+)} \Omega_{\beta}^{(-)}) =
  \frac{2 \cos \beta\pi}{\pi (1-2\beta)} =
  \frac{2(-1)^{\beta}}{\pi(1-2\beta)}$. 

\subsection{Matching process and temperature}
As easily seen from Eqs. (\ref{z1sol}) and (\ref{solasym}), by matching
these solutions, we obtain the following relations, 
\begin{eqnarray}
  \label{match1}
  B_{\rm in} &=& \frac{1}{2}\left[\frac{r_{h}^{1-\beta}}{\Omega_{\beta}^{(+)}} \left(C_{\rm in}
  \frac{\Gamma(2\beta-1)\Gamma(1-2\alpha)}{\Gamma(-\alpha_{+}+\beta)\Gamma(-\alpha_{-}+\beta)}
  + C_{\rm out}
  \frac{\Gamma(2\beta-1)\Gamma(1+2\alpha)}{\Gamma(\alpha_{-}+\beta)\Gamma(\alpha_{+}+\beta)}\right)\right.\nonumber \\
&-& \left.i\frac{r_{h}^{\beta}}{\Omega_{\beta}^{(-)}}\left(C_{\rm in}
  \frac{\Gamma(1-2\alpha){\cal D}_{\nu}^{\rm
  in}}{\Gamma(1-\alpha_{+}-\beta)\Gamma(1-\alpha_{-}-\beta)} + C_{\rm
  out}\frac{\Gamma(1+2\alpha){\cal D}_{\nu}^{\rm
  out}}{\Gamma(1+\alpha_{+}-\beta)\Gamma(1+\alpha_{-}-\beta)}\right)\right],\nonumber
  \\
  B_{\rm out} &=& \frac{1}{2}\left[\frac{r_{h}^{1-\beta}}{\Omega_{\beta}^{(+)}} \left(C_{\rm in}
  \frac{\Gamma(2\beta-1)\Gamma(1-2\alpha)}{\Gamma(-\alpha_{+}+\beta)\Gamma(-\alpha_{-}+\beta)}
  + C_{\rm out}
  \frac{\Gamma(2\beta-1)\Gamma(1+2\alpha)}{\Gamma(\alpha_{-}+\beta)\Gamma(\alpha_{+}+\beta)}\right)\right.\nonumber \\
&+& \left.i\frac{r_{h}^{\beta}}{\Omega_{\beta}^{(-)}}\left(C_{\rm in}
  \frac{\Gamma(1-2\alpha){\cal D}_{\nu}^{\rm
  in}}{\Gamma(1-\alpha_{+}-\beta)\Gamma(1-\alpha_{-}-\beta)} + C_{\rm
  out}\frac{\Gamma(1+2\alpha){\cal D}_{\nu}^{\rm
  out}}{\Gamma(1+\alpha_{+}-\beta)\Gamma(1+\alpha_{-}-\beta)}\right)\right].
\end{eqnarray}

At this stage, we impose a boundary condition $C_{\rm out}=0$, which
implies that there are no outgoing modes of scalar fields since all
modes of fields are absorbed into the black hole horizon. 
Defining a flux as
\begin{equation}
  \label{fluxdef}
  {\cal F} = \frac{2\pi}{i}U(r)\left(R^{\ast}(r)\partial_{r}R(r) - R(r)\partial_{r}R^{\ast}(r)\right),
\end{equation}
we obtain them near the geodesic orbit $r_{g}$ as
\begin{eqnarray}
  \label{fluxesss}
  & & {\cal F}_{r_{g}}^{\rm out} = - 8 |B_{\rm out}|^2 (-1)^{\beta}
  r_{g}^{-2}, \nonumber \\
  & & {\cal F}_{r_{g}}^{\rm in} = 8 |B_{\rm in}|^2 (-1)^{\beta}
  r_{g}^{-2}.
\end{eqnarray}
The reflection coefficient near the geodesic orbit $r_{g}$ is defined
as a ratio between ingoing and outgoing fluxes, and consequently it
gives
\begin{equation}
  \label{reflec}
  {\cal R_{\rm BH}} = \left|\frac{{\cal F}_{r_{g}}^{\rm in}}{{\cal
  F}_{r_{g}}^{\rm out}}\right|^2 = \left|\frac{B_{\rm in}}{B_{\rm out}}\right|^2.
\end{equation}
Furthermore, statistical-thermodynamic temperature is defined as a vacuum
expectation value of a number operator $N$ and it is related to the
reflection coefficient \cite{ko,gl},
\begin{equation}
  \label{exp}
  <0|N|0> = \frac{1}{e^{{\omega}/{T_{stat}}} - 1} = \frac{{\cal R}_{\rm
  BH}}{1-{\cal R}_{\rm BH}}.
\end{equation}
Therefore, we have a thermal temperature of black holes,
\begin{equation}
  \label{temp}
  T_{stat} = - \frac{\omega}{\ln {\cal R}_{\rm BH}}.
\end{equation}
From Eq. (\ref{match1}) with the boundary condition $C_{\rm
  out}=0$, the logarithm term in Eq. (\ref{temp}) becomes
\begin{equation}
  \label{log}
  \ln {\cal R}_{\rm BH} = \ln \left(\frac{1-i\xi}{1+i\xi}\right)= - 2i\xi +
  {\cal O}(\xi^2),
\end{equation}
where 
\begin{eqnarray}
  \label{xi}
  \xi &\equiv&
  \frac{1}{2\pi}r_{h}(-1)^{\beta+1} \Gamma(2\beta)
  \sin\pi\alpha_{+}\sin\pi\alpha_{-}({\cal D}_{\nu}^{in} - {\cal
  D}_{\nu}^{\ast in})\nonumber \\
  &\times&
  \left(\frac{r_{h}^{2-2\beta}(\frac{1}{2}p)^{1-2\beta}\Gamma^2\left(\beta+\frac{1}{2}\right)\Gamma^2(2\beta-1)}{\Gamma(\alpha_{+}+\beta)\Gamma(-\alpha_{+}+\beta)\Gamma(\alpha_{-}+\beta)\Gamma(-\alpha_{-}+\beta)}
\right. \nonumber \\
 &+&\left.\frac{r_{h}^{2\beta}(\frac{1}{2}p)^{2\beta-1}\Gamma^2\left(\frac{3}{2}-\beta\right)|{\cal
  D}_{\nu}^{\rm
  in}|^2}{\Gamma(1+\alpha_{+}-\beta)\Gamma(1+\alpha_{-}-\beta)\Gamma(1-\alpha_{+}-\beta)\Gamma(1-\alpha_{-}-\beta)}
  \right)^{-1}.
\end{eqnarray}

For simplicity, considering the main contribution of
$s$-modes(vanishing angular potential) when
$\beta=1$ as shown in Ref. \cite{cl,fro}, Eq. (\ref{xi}) is simplified
as $\xi = -\pi^2 i \omega r_{h} /3$ by using the Schwarz's
inequality. So, we obtain the ``minimal'' thermal temperature,
\begin{equation}
  \label{temp2}
  T_{stat}^{\rm min} = T_{\rm BH} = \frac{1}{2\pi r_{h}} \left(\frac{3}{\pi}\right) \approx
  \frac{1}{2\pi r_{h}},
\end{equation}
where it agrees with the well-known Hawking temperature
for the SS black hole. Therefore, we have shown that the desired
Hawking temperature can be reproduced as a ``minimal'' statistical
temperature in attraction dominant region (Region (I) and (II)).

\section{Low Energy Perturbation in the De-Sitter Region \hfil{}}
\subsection{Solution near the de Sitter Spacetimes (Region (III) and (IV))}
Now we study the solution in the expansion dominant region (region
(III) and (IV)). As $r \rightarrow r_{c}$, Eq. (\ref{radial}) becomes 
\begin{equation}
  \label{desitt}
  \left(r_{c}^2 - r^2 \right) \partial_{r}^2 R(r) + \left(\frac{2}{r}
  - \frac{4r}{r_{c}^2} \right) \partial_{r} R(r) +
\left(\frac{\omega^2 r_{c}^2}{(r_{c}^2 - r^2)} -
  \frac{\nu(\nu+1)}{r^2} - m^2\right) R(r) =0,
\end{equation}
where $r_{c}$ is nearly $(3\Lambda^{-1})^{1/2}$ as $r$ goes to
$r_{c}$. By choosing a coordinate as $y=1-r^2/r_{c}^2$ where $y$
spans from $0$ to $1$, the radial equation (\ref{desitt}) is written
as 
\begin{equation}
  \label{zeqnds}
  y(1-y)\partial_{y}^2 R(y) + \frac{1}{2}(2-5y)\partial_{y}R(y) +
  \left(\frac{\omega^2 r_{c}^2}{4y} - \frac{\nu(\nu+1)}{4(1-y)} -
  \frac{m^2 r_{c}^2}{4} \right)R(y) = 0.
\end{equation}
Singular points at $y=0$ and $y=1$ can be eliminated by choosing $R(y)
= y^{\tilde \alpha}(1-y)^{\tilde \beta} u(y)$, and Eq. (\ref{zeqnds})
is written as
\begin{eqnarray}
  \label{yint}
    & &y(1-y)\partial_{y}^2 u(y) + \left(1+2{\tilde\alpha} -
  \frac{1}{2}\left(4{\tilde\alpha} + 4{\tilde\beta} +5\right)y\right)
  \partial_{y} u(y) \nonumber \\
 & &\qquad\qquad\qquad\qquad+ \frac{1}{y}\left({\tilde\alpha}^2 +
  \frac{\omega^2r_{c}^2}{4}\right)u(y) +
  \frac{1}{1-y}\left({\tilde\beta}^2 + \frac{{\tilde\beta}}{2} -
  \frac{\nu(\nu+1)}{4}\right) \nonumber \\
 & &\qquad\qquad\qquad\qquad\qquad\qquad\qquad\qquad- \left[({\tilde\alpha} +
  {\tilde\beta})({\tilde\alpha} + {\tilde\beta}+\frac{3}{2}) +
  \frac{m^2 r_{c}^2}{4}\right]u(y) = 0.
\end{eqnarray}
Determining ${\tilde\alpha}=i\omega r_{c}/2$ and ${\tilde\beta} =
\nu/2 = -(1-\beta)/2$, Eq. (\ref{yint}) becomes
\begin{eqnarray}
  \label{eqnds}
    & &y(1-y)\partial_{y}^2 u(y) + \left(1+2{\tilde\alpha} -
  \frac{1}{2}\left(4{\tilde\alpha} + 4{\tilde\beta} +5\right)y\right)
  \partial_{y} u(y) \nonumber \\
 & &\qquad\qquad\qquad\qquad\qquad\qquad\qquad\qquad- \left[({\tilde\alpha} +
  {\tilde\beta})({\tilde\alpha} + {\tilde\beta}+\frac{3}{2}) +
  \frac{m^2 r_{c}^2}{4}\right]u(y) = 0.
\end{eqnarray}
By defining $\gamma_{\pm}\equiv (3\pm\sqrt{9-4m^2 r_{c}^2})/4$, the
solution is given as
\begin{eqnarray}
  \label{solds}
  R(y) &=& A_{\rm in} y^{\tilde\alpha}(1-y)^{\tilde\beta} F(\gamma_{-} +
  {\tilde\alpha} + {\tilde\beta},
  \gamma_{+}+{\tilde\alpha}+{\tilde\beta},
  1+2{\tilde\alpha};y)\nonumber \\
  & &\qquad\qquad + A_{\rm out} y^{-\tilde\alpha}(1-y)^{\tilde\beta} F(\gamma_{-} -
  {\tilde\alpha} + {\tilde\beta},
  \gamma_{+}-{\tilde\alpha}+{\tilde\beta},
  1-2{\tilde\alpha};y),
\end{eqnarray}
where $\gamma_{+}+\gamma_{-} =3/2$. Note that the solution
(\ref{solds}) follows a $z\rightarrow 1-z$ transformation given by
\begin{eqnarray}
  \label{z1ztransds}
  F(a,b,c;z) &=& \frac{\Gamma(a)\Gamma(c-a-b)}{\Gamma(c-a)\Gamma(c-b)}
  F(a,b,a+b-c+1;1-z)\nonumber \\
  &+& (1-z)^{c-a-b} \frac{\Gamma(c)\Gamma(a+b-c)}{\Gamma(a)\Gamma(b)} F(c-a,c-b,c-a-b+1;1-z).
\end{eqnarray}
Therefore, for the limit of $y\rightarrow 1$ ($r\rightarrow r_{g} << r_{c}$), we obtain the solution in region (III)
of Eq.(\ref{solds}),
\begin{eqnarray}
  \label{regiiisol}
  & &R_{y\rightarrow1} (r) \nonumber \\
  &=& \left(\frac{A_{\rm in}\Gamma(1+2{\tilde\alpha})\Gamma\left(-2{\tilde\beta}-\frac{1}{2}\right)}{\Gamma(1+{\tilde\alpha}-{\tilde\beta}-\gamma_{-})\Gamma(1+{\tilde\alpha}-{\tilde\beta}-\gamma_{+})}
  + \frac{A_{\rm out}\Gamma(1-2{\tilde\alpha})\Gamma\left(-2{\tilde\beta}-\frac{1}{2}\right)}{\Gamma(1-{\tilde\alpha}-{\tilde\beta}-\gamma_{-})\Gamma(1-{\tilde\alpha}-{\tilde\beta}-\gamma_{+})}\right)\left(\frac{r_{c}}{r}\right)^{1-\beta}\nonumber \\
 &+&\left(\frac{A_{\rm in}\Gamma(1+2{\tilde\alpha})\Gamma\left(2{\tilde\beta}+\frac{1}{2}\right)}{\Gamma({\tilde\alpha}+{\tilde\beta}+\gamma_{-})\Gamma({\tilde\alpha}+{\tilde\beta}+\gamma_{+})}
  + \frac{A_{\rm out}\Gamma(1-2{\tilde\alpha})\Gamma\left(2{\tilde\beta}+\frac{1}{2}\right)}{\Gamma(-{\tilde\alpha}+{\tilde\beta}+\gamma_{-})\Gamma(-{\tilde\alpha}+{\tilde\beta}+\gamma_{+})}\right)\left(\frac{r_{c}}{r}\right)^{\beta}.
\end{eqnarray}

\subsection{Asymptotic solution at region (III)}
On the other hand, for the asymptotic limit $r\rightarrow r_{g}$, the
equation of motion (\ref{eqnds}) can be written as the same form with
Eq. (\ref{asymss}) and the solution is given by Bessel functions with
different amplitudes, $D_{1}$ and $D_{2}$,
\begin{equation}
  \label{bessel2}
  R_{r_{g}}^{(y)} (r) = \frac{D_{1}}{\sqrt{r}}
  J_{-\frac{1}{2}(1-2\beta)}(pr) + \frac{D_{2}}{\sqrt{r}}
  J_{\frac{1}{2}(1-2\beta)}(pr),
\end{equation}
where $p=\sqrt{\omega^2 - m^2}$ and $\beta=\nu+1$. 
In the same way for
the previous calculation, by defining new coefficients $D_{\rm in}$ and
$D_{\rm out}$ as
\begin{eqnarray}
  \label{decomp2}
  & &D_{1} = D_{\rm in} + D_{\rm out}\nonumber \\
  & &D_{2} = i(D_{\rm in} - D_{\rm out}),
\end{eqnarray}
the solution (\ref{bessel2}) can be written as
\begin{equation}
  \label{asymds}
  R_{r_{g}}^{(y)} (r) = D_{\rm in} \left( \Omega_{\beta}^{(+)}
  \frac{1}{r^{1-\beta}} + i\Omega_{\beta}^{(-)}
  \frac{1}{r^{\beta}}\right) + D_{\rm out} \left(\Omega_{\beta}^{(+)}
  \frac{1}{r^{1-\beta}} - i\Omega_{\beta}^{(-)} \frac{1}{r^{\beta}}\right).
\end{equation}

\subsection{Matching process and temperature}
Comparing Eq. (\ref{regiiisol}) with Eq. (\ref{asymds}) to perform a
coefficient match, we obtain the relations between coefficients at region
(III) and (IV),
\begin{eqnarray}
  \label{matching2}
  D_{\rm in}&=& \frac{1}{2\Omega_{\beta}^{(+)}} \left(\frac{A_{\rm in}r_{c}^{\beta-1}\Gamma(1+2{\tilde\alpha})\Gamma(-2{\tilde\beta}-\frac{1}{2})}{\Gamma(1+{\tilde\alpha}-{\tilde\beta}-\gamma_{-})\Gamma(1+{\tilde\alpha}-{\tilde\beta}-\gamma_{+})}
  + \frac{A_{\rm out}r_{c}^{\beta-1}\Gamma(1-2{\tilde\alpha})\Gamma(-2{\tilde\beta}-\frac{1}{2})}{\Gamma(1-{\tilde\alpha}-{\tilde\beta}-\gamma_{-})\Gamma(1-{\tilde\alpha}-{\tilde\beta}-\gamma_{+})}\right)\nonumber \\
 &-& i\frac{1}{2\Omega_{\beta}^{(-)}} \left(\frac{A_{\rm in}r_{c}^{-\beta}\Gamma(1+2{\tilde\alpha})\Gamma(2{\tilde\beta}+\frac{1}{2})}{\Gamma({\tilde\alpha}+{\tilde\beta}+\gamma_{-})\Gamma({\tilde\alpha}+{\tilde\beta}+\gamma_{+})}
  + \frac{A_{\rm out}r_{c}^{-\beta}\Gamma(1-2{\tilde\alpha})\Gamma(2{\tilde\beta}+\frac{1}{2})}{\Gamma(-{\tilde\alpha}+{\tilde\beta}+\gamma_{-})\Gamma(-{\tilde\alpha}+{\tilde\beta}+\gamma_{+})}\right)\nonumber \\
D_{\rm out} &=& \frac{1}{2\Omega_{\beta}^{(+)}} \left(\frac{A_{\rm in}r_{c}^{\beta-1}\Gamma(1+2{\tilde\alpha})\Gamma(-2{\tilde\beta}-\frac{1}{2})}{\Gamma(1+{\tilde\alpha}-{\tilde\beta}-\gamma_{-})\Gamma(1+{\tilde\alpha}-{\tilde\beta}-\gamma_{+})}
  + \frac{A_{\rm out}r_{c}^{\beta-1}\Gamma(1-2{\tilde\alpha})\Gamma(-2{\tilde\beta}-\frac{1}{2})}{\Gamma(1-{\tilde\alpha}-{\tilde\beta}-\gamma_{-})\Gamma(1-{\tilde\alpha}-{\tilde\beta}-\gamma_{+})}\right)\nonumber \\
 &+& i\frac{1}{2\Omega_{\beta}^{(-)}} \left(\frac{A_{\rm in}r_{c}^{-\beta}\Gamma(1+2{\tilde\alpha})\Gamma(2{\tilde\beta}+\frac{1}{2})}{\Gamma({\tilde\alpha}+{\tilde\beta}+\gamma_{-})\Gamma({\tilde\alpha}+{\tilde\beta}+\gamma_{+})}
  + \frac{A_{\rm out}r_{c}^{-\beta}\Gamma(1-2{\tilde\alpha})\Gamma(2{\tilde\beta}+\frac{1}{2})}{\Gamma(-{\tilde\alpha}+{\tilde\beta}+\gamma_{-})\Gamma(-{\tilde\alpha}+{\tilde\beta}+\gamma_{+})}\right).
\end{eqnarray}

Now we impose a boundary condition $A_{\rm out}=0$, which means that
there are no wave modes falling into the cosmological
horizon, and we define a reflection coefficient at the geodesic orbit
$r_{g}$ as 
\begin{equation}
  \label{refl2}
  {\cal R}_{\rm DS} = \left|\frac{\tilde{\cal F}_{\rm out}}{\tilde{\cal
  F}_{\rm in}}\right| = \left|\frac{D_{\rm out}}{D_{\rm in}}\right|^2,
\end{equation}
where the ingoing and outgoing fluxes at $r_{g}$ are obtained by the
definition (\ref{fluxdef}),
\begin{eqnarray}
  \label{flux2}
  & &\tilde{\cal F}_{r_{g}}^{\rm in} = 8|D_{\rm in}|^2
  (-1)^{\beta}r_{g}^{-2}\nonumber \\
  & &\tilde{\cal F}_{r_{g}}^{\rm out} = -8|D_{\rm out}|^2
  (-1)^{\beta}r_{g}^{-2}.
\end{eqnarray}

With the help of Eq. (\ref{matching2}), the reflection
coefficient (\ref{refl2}) is written as
\begin{equation}
  \label{reflec2}
  {\cal R}_{\rm DS} = \frac{1+i{\tilde \xi}}{1-i{\tilde \xi}},
\end{equation}
where
\begin{eqnarray}
  \label{xi2}
  {\tilde \xi} &\equiv& r_{c}\sin2{\tilde\alpha}\pi
  \sin^2\frac{\pi\beta}{2}\nonumber \\
  &\times&
  \left(\frac{\pi^2r_{c}^{2(1-\beta)}(\frac{1}{2}p)^{1-2\beta}}{\Gamma(1-{\tilde\alpha}-{\tilde\beta}-\gamma_{-})\Gamma(1+{\tilde\alpha}-{\tilde\beta}-\gamma_{-})\Gamma(1-{\tilde\alpha}-{\tilde\beta}-\gamma_{+})\Gamma(1+{\tilde\alpha}-{\tilde\beta}-\gamma_{+})}\right.\nonumber
  \\
 &+&\left. \frac{r_{c}^{2\beta}\pi^2(\frac{1}{2}p)^{-(1-2\beta)}}{\Gamma(\gamma_{-}-{\tilde\alpha}+{\tilde\beta})\Gamma(\gamma_{-}+{\tilde\alpha}+{\tilde\beta})\Gamma(\gamma_{+}-{\tilde\alpha}+{\tilde\beta})\Gamma(\gamma_{+}+{\tilde\alpha}+{\tilde\beta})}\right)^{-1}.
\end{eqnarray}
By using the relations ${\tilde\beta} = -(1-\beta)/2$ and
$\Gamma(s)\Gamma(1-s) = {\pi}/{\sin\pi s}$, and the thermal temperature is given as,
\begin{equation}
  \label{temp2}
  T_{stat} = -\frac{\omega}{\ln {\cal R}_{\rm DS}} \approx - \frac{\omega}{2i{\tilde\xi}}
\end{equation}
in a leading order of $\omega$. To be explicit, for $\nu =0$,
i.e. ${\tilde \beta} =0$, Eq. (\ref{xi2}) is simplified as
\begin{equation}
  \label{simplexi}
  {\tilde\xi}=\frac{\sin2\pi{\tilde\alpha}}{\sqrt{\cos^2\pi(\gamma_{+}-\gamma_{-})- \sin^2 2\pi{\tilde\alpha}}}.
\end{equation}
Expanding Eq. (\ref{simplexi}) in a low energy limit, ${\tilde\xi} = 2 \pi
{\tilde\alpha} + {\cal O}(m^2, {\tilde\alpha}^2)$ and the ``minimal'' thermal
temperature can be obtained as
\begin{equation}
  \label{themaltempds}
  T_{stat}^{min} = T_{\rm DS} \approx \frac{1}{2\pi r_{c}},
\end{equation}
which agrees with the thermal temperature of the dS spacetimes.

\section{Discussions}
In summary, we have studied the scattering problem of the massive
scalar field on the SSdS background by assuming the low energy of the
probing field. For the small static SSdS black hole, the whole
spacetimes can be considered as a patch of two different regions (SS
black hole and dS spacetimes) and the geodesic orbit where the black
hole attraction and cosmic expansion balance out is introduced. Using
the low energy approximation, the corresponding statistical
temperatures from two horizons are derived, and it is
interesting to note that the minimal bounds of these temperatures are
equivalent to the thermal temperatures obtained from the metric.  
 
As a comment, we should perform a
fine-tuning process between the wave functions at the region (II) and
(III) in Sec. II. The wave functions of the massive scalar field should be
connected smoothly at the geodesic orbit since it is regarded as a
continuous function in the whole regions. Therefore, from
Eqs. (\ref{solasym}) and (\ref{asymds}), we define $B_{\rm in} \equiv
N_{\rm in} D_{\rm in}$ and $B_{\rm out}\equiv N_{\rm out} D_{\rm
  out}$, respectively, where $N_{\rm in}$ and $N_{\rm out}$ are
proportional constants. And then, Eqs. (\ref{reflec}) and
(\ref{refl2}) produce a relation $R_{\rm DS}R_{\rm BH} = |{\cal
  Q}|^{-2}$ by defining $|{\cal Q}|^2 = |N_{\rm in}|^2/|N_{\rm
  out}|^2$ and it determines, 
\begin{equation}
\label{nrch}
|{\cal Q}|^2 = e^{\omega\left(T_{BH}^{-1}+ T_{DS}^{-1}\right)},
\end{equation}
which is a fine-tuning for mismatched wave functions at the geodesic
orbit and it correlates the cosmological horizon $r_{c}$ with the
black hole horizon $r_{h}$. By means of this final coefficient-tuning, the
wave functions of probing fields are continuous and connected smoothly
in the whole area.

Now let us discuss about the qualitative behaviors of non-degenerate
and nearly degenerate cases of SSdS black holes. As mentioned above,
our whole calculations have been done by assuming that the
black hole horizon size is much smaller than that of the cosmological horizon,
which describes a small SSdS black hole. Since the black hole will be
hotter than the cosmological horizon, the one will evaporate faster than
the other, and the black hole horizon will disappear at the end of
evaporation. And then the SSdS spacetimes may become the pure dS spacetimes
at the end of the evaporation. 
At this stage,
one may ask a question : {\it``After evaporation at final stage, is it
  really possible to transit from the SSdS spacetimes to the dS
  ones?''} In fact, to
answer this question explicitly, one should study this model with the
gravitational back reaction by using the time evolution of the small
SSdS black hole, however we do not discuss anymore because of
complexity of this analysis.

On the other hand, we naturally consider the degenerate case of
horizons that the black hole horizon is equal to the cosmological
horizon. In this case, we expect that the black hole will be in the
thermal equilibrium state since the black hole and
cosmological horizons have the same surface gravity. Therefore, when the
black hole horizon ($r_{h}$) approaches to the cosmological horizon
($r_{c}$), the two $S^{2}$ hypersufaces ($r_{h}$ and $r_{c}$) meet
each other in a hypersurface ($r_{g}$), and this is somewhat trivial
to discuss. To investigate the small perturbation from the thermal
equilibrium, we can consider the nearly degenerate case whose metric
is given by  
\begin{equation}
  \label{neardg}
  (ds)^2 = -\frac{1}{\sqrt{\Lambda}} \left(1+\frac{2}{3} \delta
  {\rm cos} R\right){\rm sin}^2R d^2T + \frac{1}{\Lambda}\left(1-\frac{2}{3}\delta
  {\rm cos} R\right)d^2R + \frac{1}{\Lambda}\left(1-2\delta {\rm cos}\right)d^2\Omega_{(2)}
\end{equation}
with defining new time and radial coordinates $T$ and $R$ by
\begin{eqnarray}
  \label{defnew}
  t&=&\frac{1}{\delta \sqrt{\Lambda}}T, \nonumber \\
  r&=&\frac{1}{\sqrt{\Lambda}} \left(1-\delta {\rm cos}R - \frac{1}{6}\delta^2\right),
\end{eqnarray}
and $9M^2\Lambda = 1 - 3 \delta^2$, where $0\le \delta \le 1$.
And the temperatures from two horizons are
given by Ref. \cite{bh1}
\begin{equation}
  \label{dgtemp}
  T_{{\rm h,c}} = \frac{1}{2\pi} \sqrt{\Lambda}\left(1\pm
  \frac{2}{3}\delta\right) + O(\delta^2),
\end{equation}
where the upper (lower) sign is for the black hole (cosmological)
horizon. Note that if $\delta$ goes to zero, Eq. (\ref{dgtemp})
describes the degenerate temperature characterized by only a
cosmological constant $\Lambda$, and the instability of the nearly
degenerate SSdS black hole has been studied by adding the one loop
quantum correction terms \cite{bh2}. As a further work, if the
scattering problem in the nearly degenerate SSdS black hole background
is solved by use of the low energy perturbation method, the results of
both temperatures will be compatible with Eq. (\ref{dgtemp}).

\vspace{1.5cm}

{\bf Acknowledgments}\\
This work has been achieved partially for participating in the
APCTP-PIMS Summer Workshop at SFU, Canada. W. T. Kim and J. J. Oh
would like to thank PIMS for hospitality during staying at SFU.
J. J. Oh and K. H. Yee would like to thank V. P. Frolov for helpful
comments while they were participating in the 6th APCTP Winter School at
POSTECH, Korea. This work was supported by the Korea Research
Foundation Grant, KRF-2001-015-DP0083.

%%%%%%%%%%%%%%%%%%%% References %%%%%%%%%%%%%%%%%%%%%%%%%

\end{document}